\documentclass[prl,twocolumn,showpacs,showkeys,preprintnumbers,amsmath,amssymb,nofootinbib]{revtex4-1}
\usepackage[utf8]{inputenc}

\usepackage{amsmath,amssymb}
\usepackage{lineno,hyperref}
\usepackage{graphicx}
\usepackage{bm}
\usepackage{bbm}
\usepackage{color}
\usepackage{amsfonts}

\modulolinenumbers[5]

\def\be{\begin{equation}}
\def\ee{\end{equation}}

	\newcommand{\ncd}{\newcommand}
	\ncd{\mrm}    {\mathrm}
	\ncd{\beq} {\begin{equation}}
	\ncd{\eeq} {\end{equation}}
	\ncd{\nn}{\nonumber}

	\def\d{{\rm d}}
	\def\D{{\rm D}}

	\def\basis[#1]{\frac{\partial}{\partial #1}}
	\def\dt[#1]{\frac{\d}{\d #1}}

	\def\ric{{\rm Ric}}
	\def\ri{{\rm i}}

\begin{document}

\title{Contact Geometry in Superconductors and {New Massive Gravity}}

\author{Daniel \surname{Flores-Alfonso}}
\email[]{daniel.flores@xanum.uam.mx}
\affiliation{Departamento de F\'isica,
Universidad Aut\'onoma Metropolitana - Iztapalapa,
Avenida San Rafael Atlixco 186, A.P. 55534, C.P. 09340, Ciudad de M\'exico, Mexico}

\author{Cesar S.~\surname{Lopez-Monsalvo}}
\email[]{cslopezmo@conacyt.mx}
\affiliation{Conacyt-Universidad Aut\'onoma Metropolitana Azcapotzalco,  Avenida San Pablo Xalpa 180, Azcapotzalco, Reynosa Tamaulipas, C.P. 02200, Ciudad de M\'exico, Mexico}

\author{Marco \surname{Maceda}}
\email[]{mmac@xanum.uam.mx}
\affiliation{Departamento de F\'isica,
Universidad Aut\'onoma Metropolitana - Iztapalapa,
Avenida San Rafael Atlixco 186, A.P. 55534, C.P. 09340, Ciudad de M\'exico, Mexico}

\begin{abstract}
The defining property of every three-dimensional $\varepsilon$-contact manifold is
shown to be equivalent to requiring the fulfillment of London's equation in 2+1 electromagnetism. To illustrate this point, we show that every such manifold that is also K-contact and $\eta$-Einstein is a vacuum solution to the most general quadratic-curvature gravity action, in particular of New Massive Gravity. As an example we analyse $S^3$ equipped with a contact structure together with an associated metric tensor such that the canonical generators of the contact distribution are null. The resulting Lorentzian metric is shown to be a vacuum solution of three-dimensional massive gravity. Moreover, by coupling the New Massive Gravity action to Maxwell-Chern-Simons we obtain a class of charged solutions stemming directly from the para-contact metric structure. Finally, we repeat the exercise for the Abelian Higgs theory.
\end{abstract}

\pacs{04.60.Kz, 11.10.Kk, 11.15.Wx}
\keywords{Contact geometry, New Massive Gravity}
\maketitle

\section{Introduction}
There is a long tradition on applications of geometrical methods in physics; over the years branches as Hamiltonian dynamics, geometric optics, fluid dynamics and General Relativity have benefited from the techniques developed with a geometric perspective~\cite{Trautman:1999ta,Robinson:1985gw,Robinson:1988,Arnol'd:1989,Robinson:1993,Ghrist:2000}. In the case of pseudo-Riemannian manifolds, this becomes even more true with the use of para-contact geometry~\cite{Sato:1976,Adati:1977,kaneyuki1985,Adati:1985,Zamkovoy:2009,Bejan:2014}. More specifically, para-Sasakian geometry has paved its way into General Relativity as a tool to analyze Ricci solitons, lightlike hypersurfaces, Killing vectors and associated horizons~\cite{Matsumoto:1977,Rahman:1995,Duggal:1996,Duggal:2010,CALVARUSO20151,Blaga:2020}. In this Letter, we discuss further implications of para-Sasakian structures within the realm of New Massive Gravity (NMG)~\cite{Bergshoeff:2009hq}.
In particular, we show that the structure gives rise to a distinguished Trkalian flow; a special type of Beltrami flow.

Beltrami fields where originally introduced in the realm of hydrodynamics to describe flows whose stream lines are parallel to their vorticity.
In the case of electromagnetism, these have received the name of \emph{force-free} magnetic fields. That is, magnetic fields such that the induced currents in a conducting medium experience a vanishing  Lorentz force.
This feature is, indeed, a property of the medium and it is described by London's constitutive relations. In this letter we show that such relations are completely captured by the geometry of  a class of three dimensional metric contact manifolds whose structural elements give rise to propagating  force-free fields. In particular, we study the case of $S^3$ endowed with a contact structure together with an associated metric such that the generators of the contact distribution together with the Reeb vector field generate a null triad for a (2+1) spacetime. We explicitly obtain the conditions for the metric to be a  solution to New Massive Gravity for the vacuum, Maxwell-Chern-Simons and Abelian-Higgs cases.

\section{Beltrami fields and superconductors}

Let us begin by considering a 3-dimensional manifold endowed with a $\varepsilon$-contact metric structure~\cite{Murcia:2019cck}. That is, a contact metric manifold such that the contact 1-form $\eta$ and the metric $g$ satisfy the relations
\begin{equation}
 \eta=\star\ell\d\eta,\quad \tilde g^{-1}(\eta,\eta)=\varepsilon\ell^{-2}, \label{forcefree}
\end{equation}
where $\star$ is the Hodge star associated with a metric $\tilde g$ in the conformal class defined by $\tilde g = \ell^2 g$ with $\ell \neq 0$\footnote{Our analysis does not depend on the particular value of $\varepsilon$, for convenience we have chosen to use $\varepsilon=1$. We specialize to a Lorentzian metric yet only equations \eqref{eq.boxeta} and \eqref{Boxl} depend on the signature of the metric.}.

To each contact 1-form $\eta$ there is a distinguished vector field $\xi$ defined by the conditions $\eta(\xi) = 1$ and $\dot\iota_\xi \d \eta = 0$. If the metric $g$ satisfies the condition $g(\xi) = \eta$, then the field $\xi$ is a Beltrami field, namely $\xi = \ell\ {\rm curl}( \xi)$.

It is straightforward to verify that
	\beq
	\label{eq.boxeta}
	(\delta\d  +1/\ell^2)\eta=0,\quad\text{and}\quad \delta \eta=0. 
	\eeq
The first equation is reminiscent of the sourceless Proca field equation,
in which case $A=\eta$. Then, the second of the two equations above is the Lorenz gauge. This interpretation means that equation \eqref{forcefree} is a sort of ``square root'' of the Proca field equation. The first time this concept was explored was in reference~\cite{Townsend:1983xs}. Therein, the concept of self-duality for gauge fields was extended to odd dimensions. In three dimensions self-duality is defined by
	\begin{equation}
 	\star F=\frac{1}{\ell} A,
	\end{equation}
but this is merely equation \eqref{forcefree} under our
gauge field interpretation. Moreover, equations \eqref{eq.boxeta} allow us to write
	\begin{equation}
 	\left(\Delta+1/\ell^2\right)\star F=0, \label{Boxl}
	\end{equation}
where $\Delta=\d\delta+\delta\d$,
which is exactly the force-free field equation
associated with superconducting media~\cite{Chandrasekhar:1957,Kholodenko}. 
The signature of the metric determines if $\Delta$ is a hyperbolic or elliptic differential operator.
Additionally, the inhomogeneous Maxwell equation  together with \eqref{forcefree} yields the relation
\begin{equation}
 \d\star F=J=\frac{1}{\ell}F, \label{Maxwell}
\end{equation}
where $J$ is the \emph{induced} current 2-form in the medium. In this sense, equation \eqref{forcefree} corresponds to a constitutive relation for a conducting medium, that is, a relation between the induced current and the field strength characterizing the response of the medium to  electromagnetic stimuli. Indeed, \eqref{forcefree} is a metric relation between the contact form and its exterior derivative for the para-Sasakian class of contact metric manifolds. 
Moreover, at this point, we are ready to rewrite equation \eqref{forcefree} as 
\begin{equation}
 \d\star J=\frac{1}{\ell^2}F, \label{London}
\end{equation}
which is no other than London's constitutive relation  for superconducting media~\cite{Sternberg:2012}. This is a remarkable result, since we have not yet used any particular form of the metric. That is, an $\varepsilon$-contact 3-manifold represents a superconducting medium.
In particular, this provides us with an interpretation of the conformal parameter $\ell$ as the penetration depth in the medium.

Finally, let us note that a contact form $\eta$ and a gauge potential $A$ are a priori very different mathematical objects which can be identified only in a particular case. The gauge potential $A$ corresponds to a connection on a principal U(1) bundle over the space-time. The  connection itself is a one-form on the total space of the principal bundle. However, it is locally represented by $\mathfrak{u}(1)$-valued one-forms. 
On overlapping coordinate charts the corresponding local one-forms must differ only by a gauge transformation. This condition need not be met by an arbitrary contact form $\eta$.

\section{ Massive Gravity: K-contact and $\eta$-Einstein structures}

A special case of $\varepsilon$-contact manifolds is given by those that are para-Sasakian.
In higher dimensions, some para-Sasakian geometries analogous to the one we study in the next section have been found to be Einstein-Gauss-Bonnet 
vacua~\cite{Bravetti_2015}. However, in three dimensions
the quadratic-curvature Gauss-Bonnet term vanishes. For this reason
we may consider instead the most general quadratic-curvature theory in three dimensions
\begin{equation}
 S[g]=\int d^3x\sqrt{-g}\left(R-2\Lambda+\beta_1R^2+\beta_2R_{\mu\nu}R^{\mu\nu}\right). \label{R2}
\end{equation}
However, Gauss-Bonnet is a ghost-free theory, thus a closer analogy between theories is provided
by New Massive Gravity, which is also 
ghost-free~\cite{Bergshoeff:2009hq}. The action is given by
\begin{equation}
 S[g]=\int d^3x\sqrt{-g}{\cal L}_{\rm{NMG}}, \label{NMG}
\end{equation}
with
\begin{equation}
 {\cal L}_{\rm{NMG}}=\frac{1}{2\kappa^2}\left[R-2\Lambda-\frac{1}{m^2}\left(|\ric|^2-\frac{3}{8}R^2\right) \right]. \label{LNMG}
\end{equation}
Here $\Lambda$ is the cosmological constant and $m$ is the mass of the propagating degrees of freedom. It is a famous result that this theory is equivalent at the linearized level to the (unitary) Fierz-Pauli action for a massive field with
spin two. The equations of motion are
\begin{equation}
 \ric-\frac{1}{2}Rg+\Lambda g-\frac{1}{2m^2}K=0 \label{EOM}
\end{equation}
where $K$ is a tensor with components

\begin{align}
K_{\mu\nu} =&~2 \Box R_{\mu\nu} - \frac 12 \left( g_{\mu\nu} \Box + \nabla_\mu \nabla_\nu - 9 R_{\mu\nu} \right) R 
\nonumber \\[4pt]
&- 8 R_{\mu\alpha}R^\alpha{}_\nu + g_{\mu\nu} \left( 3 R^{\alpha\beta} R_{\alpha\beta} - \frac {13}8 R^2 \right),
\label{Ktensor}
\end{align}
where $\Box$ is defined by the trace of the second covariant derivative, e.g., $\Box\ric=\nabla_{\alpha}\nabla^{\alpha}\ric$.

A special class of contact metric manifolds are those for which
\begin{equation}
 \ric=\lambda g+\omega\eta\otimes\eta,
\end{equation}
where $\lambda$ and $\omega$ are constants~\cite{okumura1962}. In such case we say that the manifold $(M,g)$ is $\eta$-Einstein. If in addition, the Reeb vector field is Killing then we say $M$ is a K-contact $\eta$-Einstein manifold.

In the following we show that every K-contact $\eta$-Einstein manifold is a vacuum solution of NMG assuming that \eqref{forcefree} holds.
From the $\varepsilon$-contact metric structure, we have $|\eta|_g^2 = \varepsilon$ and $\eta(\xi) = 1$, where $\xi$ is the associated Reeb vector. It follows that $\eta_\mu \xi^\mu = 1 = \varepsilon \eta_\mu \eta_ \nu g^{\mu\nu}$; in consequence $\xi^\mu = \varepsilon g^{\mu\nu} \eta_\nu =: \varepsilon \eta^\mu$. Considering now that \eqref{forcefree} implies
\be
\d\eta = \ell^{-1} s_g \star \eta, \label{starsg}
\ee
we obtain 
\be
\nabla_\mu \eta_\nu - \nabla_\nu \eta_\mu = \ell^{-1} s_g \epsilon_{\mu\nu\sigma} \eta^\sigma,
\label{equeta}
\ee
where $\epsilon_{\mu\nu\sigma} := \sqrt{|g|}\varepsilon_{\mu\nu\sigma}$ is the Levi-Civita tensor density with $\varepsilon_{012} = +1$. Now, from $\varepsilon g = \eta \otimes \eta + s_g d\eta (\phi \otimes 1)$, we have $\varepsilon \mathcal L_\xi g = \mathcal L_\xi \eta \otimes \eta + \eta \otimes \mathcal L_\xi \eta + s_g \mathcal L_\xi d\eta (\phi \otimes 1)$, where $\mathcal L$ denotes the Lie derivative. Then, by using the property that the Lie derivative and the exterior derivative $d$ commute when acting on $p$-forms, together with $\mathcal L_\xi \eta = 0$, we obtain $\mathcal L_\xi g = 0$. This result shows that $\xi$, or equivalently $\eta$, is Killing\footnote{We have a K-contact structure.}.

With the Killing equation $\nabla_\mu \eta_\nu + \nabla_\nu \eta_\mu = 0$ at our disposal, Eq.~(\ref{equeta}) becomes 
\be
\nabla_\mu \eta_\nu = \frac 12 \ell^{-1} s_g \epsilon_{\mu\nu\sigma} \eta^\sigma.
\ee
It follows that 
\begin{align}
\Box\eta=-\frac{1}{2}\ell^{-2}s_g\eta,
\end{align}
and
\begin{align}
(\nabla_\alpha \eta_\mu) (\nabla^\alpha \eta_\nu) =&~\frac 14 \ell^{-2} s_g^2 \epsilon_{\alpha\mu\sigma} \epsilon^\alpha{}_{\nu\rho} \eta^\sigma \eta^\rho 
\nonumber \\[4pt]
=& \frac 14 \ell^{-2} s_g (\varepsilon \ell^{-2} g_{\mu\nu} - \eta_\mu \eta_\nu).
\label{DetaDeta}
\end{align}
Between equations \eqref{starsg} and \eqref{DetaDeta} we have used several key identities from~\cite{Murcia:2019cck}.

Notice that for a 0-form $f$ we have $\Box f=-\Delta f$. However, for higher degree differential forms the Laplace-de Rham and the Laplace-Beltrami differential operators are related by the Weitzenb\"ock identity~\cite{Choquet-Bruhat,Barrientos:2019msu}. In our case
\begin{equation}
 \Box\eta+\Delta\eta=\frac{1}{2}\ell^{-2}s_g\eta,
\end{equation}
which can be calculated directly cf.~\cite{Barrientos:2019msu}.

A straightforward calculation gives
\begin{align}
\Box\ric=\frac 12 \omega \ell^{-2} s_g \left( \varepsilon \ell^{-2} g - 3 \eta\otimes \eta \right).
\label{boxricci}
\end{align}
It is important to stress the decomposition of this quantity in a structure similar to that of the Ricci tensor.

Using Eq.~(\ref{boxricci}) in the field equations, we obtain the result that the $\varepsilon$-contact structure is a solution provided that
\begin{align}
\ell^2 \Lambda =&~\frac14 \frac {12 s_g \lambda + 11 \lambda^2 + 4 s_g \varepsilon \omega + 18 \varepsilon \lambda \omega + 3\omega^2}{6 s_g + 5 \lambda + 7 \varepsilon \omega}, \label{llambdam1}\\[4pt]
\ell^2 m^2 =& -\frac 14 (6 s_g + 5 \lambda + 7 \varepsilon \omega).
\label{llambdam2}
\end{align}

Notice that the NMG coupling constants satisfy the following relation
\begin{equation}
m^2 =- \frac {(6 s_g + 5 \lambda + 7 \varepsilon \omega)^2}{12 s_g \lambda + 11 \lambda^2 + 4 s_g \varepsilon \omega + 18 \varepsilon \lambda \omega + 3\omega^2}\Lambda, 
\label{lambdam}
\end{equation}
independently of the value of the conformal parameter $\ell$.


\section{Novel configurations of NMG from a para-contact metric structure}

Let us consider a topological three-dimensional sphere
endowed with its standard contact structure, parametrized by the one-form
\begin{equation}
 \eta=\frac{1}{2}\left(\d\psi+\cos\theta\d\phi\right), \label{eta}
\end{equation}
where we have used Euler angles $0\leq\psi\leq 4\pi$, $0\leq\theta\leq \pi$ and 
$0\leq\phi\leq 2\pi$ to coordinate the manifold. 
Notice that for the contact sphere $w=\psi/2$, $q=\phi/2$ and $p=-\cos\theta$ are the set of local coordinates in the Darboux theorem~\cite{Blair}. This is not surprising when one
keeps in mind the Hopf fibration of the hypersphere. 
A contact form $\eta$ defines a
unique vector field $\xi$ satisfying the conditions $\dot\iota_\xi \d \eta=0$ and $\dot\iota_\xi \eta=1$  known as the \emph{Reeb} vector field. Indeed, in the case of $S^3$  equipped with the contact form \eqref{eta} the Reeb field is tangent to the
Hopf circle fiber. 

A three-dimensional contact structure is a completely non-integrable distribution of two-dimensional planes in the tangent bundle. The generators of the contact distribution are vector fields annihilated by the contact 1-form. Thus, in the present case, these are given by
\begin{equation}
 Q=2\left(\frac{\partial}{\partial \phi}
 -\cos\theta\frac{\partial}{\partial \psi}\right),
 \quad \text{and} \quad
 P=\frac{1}{\sin\theta}\frac{\partial}{\partial \theta}, \label{PQ}
\end{equation}
which, in particular, show that the contact distribution given by \eqref{eta}
is bracket-generating~\cite{Kobayashi} as we have
\begin{equation}
 [P,Q]=\xi,\quad [\xi,Q]=0, \quad \text{and} \quad 
 [\xi,P]=0. \label{heis}
\end{equation}
This is to say, $P$ and $Q$ together with their iterated Lie brackets generate
a basis for the tangent bundle.
Moreover, equation \eqref{heis} exhibits the fact that the non-coordinate basis $\{P,Q,\xi\}$ satisfies the Heisenberg algebra.

There is a certain freedom in choosing a metric associated with a contact structure~\cite{Lopez-Monsalvo:2020}. In the present work, we consider a metric satisfying
\beq
\label{eq.cond1}
g(\xi,\xi) = 1 \quad \text{and} \quad g(\xi,P) = g(\xi,Q) =0,
\eeq
together with
\beq
\label{eq.cond2}
g(Q,Q) = g(P,P) = 0.
\eeq
Conditions \eqref{eq.cond1} merely state that the metric is compatible with the contact 1-form, that is, the Reeb vector field is normalized and is orthogonal to the generators of the contact distribution; conditions \eqref{eq.cond2} imply that the metric is Lorentzian.  These are the defining properties of an associated metric to the almost para-contact structure
\begin{equation}
 \varphi(\xi)=0, \quad \varphi(Q)=Q, \quad \text{and} \quad \varphi(P)=-P, \label{varphi} 
\end{equation}
representing a reflection in the $\theta$ direction of the contact distribution. Therefore, $(S^3,\eta,\xi,\varphi,g)$, is a para-contact manifold. Here,  the metric $g = \eta \otimes \eta - \d \eta \circ (\varphi \otimes \mathbbm{1})$
on the contact sphere,  whose line element in local coordinates is given by
\begin{equation}
 \d s^2=\frac{1}{4}\left(\d\psi^2+2\cos\theta\d\psi\d\phi-
 4\sin\theta\d\theta\d\phi+\cos^2\theta\d\phi^2\right), \label{metric}
\end{equation}
defines a (2+1) spacetime where $(P,Q,\xi)$ is its Newmann-Penrose null triad. Moreover, the congruences associated with the null vector fields $P$ and $Q$ are geodesics and  have no expansion, shear nor twist, therefore, $g$ defines a Kundt spacetime. Notice that $\xi$ corresponds to the spacelike vector field $m$ whilst $P$ and $Q$ to the null vector fields $l$ and $n$, respectively. From this,
it is straightforward to construct the orthonormal triad
\begin{align}
 \sqrt{2}e_0&=P+Q=
 -2\cos\theta\frac{\partial}{\partial \psi}
 +\frac{1}{\sin\theta}\frac{\partial}{\partial \theta}
 +2\frac{\partial}{\partial \phi}, \\
 \sqrt{2}e_1&=P-Q=
 2\cos\theta\frac{\partial}{\partial \psi}
 +\frac{1}{\sin\theta}\frac{\partial}{\partial \theta}
 -2\frac{\partial}{\partial \phi}, \\
 e_2&=\xi=2\frac{\partial}{\partial \psi},
\end{align}
so that, using the coframe \{$e^0,e^1,e^2$\}, the line element of the geometry is expressed as $\d s^2=-e^0e^0+e^1e^1+e^2e^2$
where the Lorentz signature is manifest. Hence, the geometry is not given by the standard round Riemannian metric. Nor is equation \eqref{metric}
the canonical Lorentz metric on the sphere.
Even though we have chosen the standard contact structure on the hypersphere.

The present para-contact sphere is not Einstein, however, 
it is $\eta$-Einstein, that is, the Ricci tensor satisfies~\cite{okumura1962}
\begin{equation}
 {\rm Ric}=\frac{1}{2}g-\eta \otimes \eta. \label{etaEin}
\end{equation}
It has been established that three-dimensional $\eta$-Einstein Sasakian manifolds must have constant sectional curvature
when restricted to planes in the contact distribution~
\cite{Boyer:2004eh}.
Furthermore, if the value of this constant is -3 the metric is nil
(a.k.a. Heisenberg)~\cite{Boyer:2004eh,Belgun:2000}.
For metric \eqref{metric} this constant is equal to 3 (when scaled for compatibility with~\cite{Boyer:2004eh}).
This suggests to us that the metric is nil and we attribute this difference in signs
to the geometry's para-Sasakian nature. 

The metric admits four solutions to the Killing vector field equations. We write them in Euler angles and in Darboux coordinates with Heisenberg basis
\begin{align}
 \xi_1&=\frac{\partial}{\partial \psi}=\frac{1}{2}\xi,\label{xi1}\\
 \xi_2&=\frac{\partial}{\partial \phi}=\frac{1}{2}Q-\frac{1}{2}p\xi,\label{xi2}\\
 \xi_3&=\phi\frac{\partial}{\partial \psi}+
 \frac{1}{\sin\theta}\frac{\partial}{\partial \theta}
 =P+q\xi,\label{xi3}\\
 \xi_4&=\phi\frac{\partial}{\partial \phi}
 +\frac{\cos\theta}{\sin\theta}\frac{\partial}{\partial \theta}
 = qQ-pP-pq\xi.\label{xi4}
\end{align}
The first three vector fields form a notable subalgebra
\begin{equation}
 [\xi_2,\xi_3]=\xi_1,\quad [\xi_1,\xi_2]=0, \quad \text{and} \quad 
 [\xi_1,\xi_3]=0, \label{Kheis}
\end{equation}
while the fourth vector field acts on the former by Lie bracket as
\begin{equation}
 [\xi_4,\xi_1]=0,\quad [\xi_4,\xi_3]=\xi_3, \quad \text{and} \quad 
 [\xi_4,\xi_2]=-\xi_2. \label{KV4}
\end{equation}
Equations \eqref{Kheis} and \eqref{KV4} imply that the geometry \eqref{metric} is one of 
only three possible Lorentzian left-invariant Heisenberg metrics~\cite{Rahmani} --- one 
of which is Minkowski spacetime.
This is, indeed, what was suggested to us above when we examined the metric's constant $\varphi$-holomorphic
sectional curvature~\cite{tanno1969}.
However, it was established in~\cite{D'Ambra} that closed simply connected
Lorentzian manifolds must have compact isometry groups. Thus, in the spirit
of~\cite{AyonBeato:2004if} we inspect the Killing vector fields
searching for incompatibilities with the defining identifications of spacetime.
From equations \eqref{xi3} and \eqref{xi4} we see that it is precisely the $\phi$-dependence of $\xi_3$ and $\xi_4$ which is incompatible with the identification
$\phi\sim\phi+2\pi$ of the three-sphere, as they would not be single valued. Hence, the only Killing fields are \eqref{xi1} and \eqref{xi2} which yield a compact isometry group, U(1)$\times$U(1),
as required. This distinction between local and global structures is analogous
to the renowned Ba\~nados-Teitelboim-Zanelli (BTZ) black hole~\cite{Banados:1992wn} which is locally diffeomorphic to Anti-de Sitter spacetime (AdS) but not globally.
The present para-contact sphere is only locally equivalent to a Lorentz-Heisenberg spacetime.

\subsection{NMG Vacuum}

Let us first consider the most general quadratic-curvature
theory \eqref{R2}. We find that the metric is a solution of the theory whenever
\begin{equation}
 \ell^2=\frac{1}{8\Lambda},\quad\text{and}\quad
 \beta_1=-\frac{1}{8\Lambda}-3\beta_2,
\end{equation}
In other words, the cosmological constant determines the characteristic length scale and the quadratic couplings are restricted.

From now on we specialize to NMG.
In particular, for $s_g = -1, \lambda = 1/2, \omega = -1, \varepsilon = +1$, associated to the metric \eqref{metric} of our para-Sasakian 3-sphere, expressions \eqref{llambdam1} and \eqref{llambdam2} become
\be
\ell^2 \Lambda = \frac 18, \qquad \ell^2 m^2 = \frac {21}8.
\label{vacuum}
\ee

One might also wonder if the metric
is a solution of the theory when the action is additionally coupled to the Cotton tensor, see for example~\cite{Garcia:2017}.
We find the answer to be positive.
Moreover, we mention that the Jordan normal form of the metric's Cotton tensor reveals the spacetime to be of ``Petrov'' type D~\cite{Garcia:2003bw,Garcia:2017}.

Now, equations \eqref{vacuum} tell us that the Heisenberg group with one of its left-invariant metrics is a solution of NMG. Moreover, we have checked that the Euclidean metric is also a solution to the equations of motion \eqref{EOM}.
This is one of the eight Thurston geometries. These geometries have been vastly studied in mathematics and physics.
In string theory, these geometries have been studied in the framework of string dualities~\cite{Gegenberg:2002xj}.Therein, the geometries have been found to be dual amongst themselves with one exception, the sol geometry. They have also been studied in string-inspired three dimensional gravity~\cite{Gegenberg:2003yz}.
In this light, it is a natural question if all 
Thurston geometries are NMG vacua~\footnote{We thank Eloy Ay\'on-Beato for
suggesting this question to us.}.
The answer is positive, all eight Thurston geometries are solutions to NMG.
Furthermore, when considering Lorentzian signature there are, instead of eight,
four relevant geometries~\cite{Dumitrescu:2010}, two of which have constant sectional curvature: Minkowski and AdS. The remaining two are the Lorentzian versions of the nil and sol geometries. We intend to report the full details
concerning these solutions in a forthcoming article.

\subsection{Coupling to the Maxwell-Chern-Simons Action}

Drawing from our previous examination of equation \eqref{forcefree} where the Hodge-star operator is associated to the metric \eqref{metric}, it is straightforward to verify that $(S^3,\eta,g)$ is an $\varepsilon$-contact structure. In addition, since every contact form on a three-dimensional  manifold represents a solution of the Maxwell's equations~\cite{Dahl:2008}, let us consider a gauge potential given by $A=-2q\eta$ so that the field strength is given by
\begin{equation}
 F=\d A=q\sin\theta\d\theta\wedge\d\phi, \label{Dirac}
\end{equation}
the standard homogeneous field strength on a two-sphere.

To understand how a field like \eqref{Dirac}
is supported by the para-Sasakian 3-sphere and what is the nature of the corresponding induced current, 
 we consider two obvious choices, namely, NMG coupled to Maxwell-Chern-Simons theory and alternatively coupled to the Abelian Higgs model.

Consider the NMG action functional \eqref{NMG} coupled to Maxwell-Chern-Simons theory (MCS)
\begin{equation}
 S[g,A]=\int \d^3x\sqrt{-g}{\cal L}_{\rm{NMG}}+S_{\rm{MCS}},
\end{equation}
where
\begin{equation}
 S_{\rm{MCS}}=\frac{1}{2}\int -F\wedge\star F+\mu A\wedge F. \label{MCS}
\end{equation}
Note that the first term of the MCS action is the helicity integral of the field, that is, a measure of the degree in which the field lines are linked~\cite{otway2015backlund}. Thus, the Maxwell equation \eqref{Maxwell} is satisfied provided $\ell=1/\mu$, yielding
\begin{equation}
 (\Delta+\mu^2)\star F=0,
\end{equation}
for equation \eqref{Boxl}. 
Notice that the Chern-Simons coupling constant $\mu$ determines
the mass of the gauge field. In the present case it also
fixes the characteristic length scale $\ell$ of spacetime.
Since the field is Trkalian then the
topologically massive gauge theory \eqref{MCS} is gauge invariant.

For this NMG-MCS theory the gauge field is supported by
the $\varepsilon$-contact provided
\begin{equation}
 q^2=\frac{\mu^2-8\Lambda}{4\kappa^2\mu^4}, \quad
 \text{and} \quad
 m^2=\frac{21\mu^4}{16(\mu^2-4\Lambda)}, \label{MCStuning}
\end{equation}
hold. Since $q^2\geq0$ this provides us with a restriction $\mu^2-8\Lambda\geq0$. When these inequalities are saturated we recover \eqref{vacuum}. Hence, this charged solution smoothly connects with the vacuum case.
\subsection{Coupling to the Abelian Higgs Model}

We now move on to the Abelian Higgs theory, which generalizes the Ginzburg-Landau theory
where superconductors were originally described by Abrikosov~\cite{Abrikosov}. We also refer the reader to~\cite{Nielsen:1973cs} for a closer analogue of the following configuration and to \cite{Canfora:2020ppn} for recent work on gravitating superconducting configurations.

We now couple the Abelian Higgs theory to New Massive Gravity
\begin{equation}
 S[g,A,\Phi]=\int \d^3x\sqrt{-g}{\cal L}_{\rm{NMG}}+S_{\rm{AH}},
\end{equation}
with
\begin{equation}
 S_{\rm{AH}}=\int -\frac{1}{2}F\wedge\star F+\frac{1}{2}\D\Phi\wedge\star\D\Phi^\dagger-\star V(|\Phi|^2). 
\end{equation}
The Higgs field $\Phi$ is in general complex-valued and its covariant derivative is given by $\D\Phi=\d\Phi-\ri A\Phi$. The Higgs field satisfies the equation of motion
\begin{equation}
 \star\D\star\D\Phi=V'. \label{Higgs}
\end{equation}
Here, we consider a contribution  from the scalar field to the energy momentum tensor, so that
\begin{equation}
 T^\Phi_{\mu\nu}=-\D_{\mu}\Phi\D_{\nu}\Phi
 +\frac{1}{2}g_{\mu\nu}\left(\D_\alpha\Phi\D^\alpha\Phi-V\right).
\end{equation}
Additionally, the Higgs field's electric current is given by
\begin{equation}
 \star J=-\ri\Phi\D\Phi^\dagger.
\end{equation}
By considering a constant real-valued Higgs field $\Phi=h$ the previous
equation becomes a London equation $\star J=h^2 A$ which when compared to
\eqref{London} shows that the value of the Higgs field plays the role of $\mu$
in the MCS case. Indeed, considering $\ell=1/h$ is sufficient for the Maxwell equations
to hold. Moreover, the Higgs equation \eqref{Higgs} fixes the self-interaction potential to
\begin{equation}
 V(|\Phi|)=\frac{\lambda}{4}|\Phi|^4.
\end{equation}
The scalar field's equation of motion also fixes the charge of the Maxwell field \eqref{Dirac} through $4q^2=\lambda$. When compared to the MCS configuration above, it possesses a rigid Maxwell field which compensates the extra degree of freedom coming from the Higgs field.

The Maxwell and Higgs fields self-gravitate on the background whenever
\begin{equation}
 h^2=\frac{1\pm\sqrt{1-32\kappa^2 \lambda\Lambda}}{2\kappa^2 \lambda}, \quad
 \text{and} \quad
 m^2=\frac{21 h^2}{8(1+2\kappa^2 \lambda h^2)}.
\end{equation}
Notice that the limit $\lambda\to0$ turns off all the field content simultaneously. Only the negative branch of $h^2$ in the previous equation
is well defined. This branch smoothly connects to the vacuum solution, given by \eqref{vacuum}.

\section{Closing remarks}

In this manuscript, we showed that the metric relation \eqref{forcefree} -- which holds for every 3-dimensional  manifold  equipped with an $\varepsilon$-contact strucutre -- serves as a constitutive relation for electromagnetic fields such that their potential 1-form is proportional to the contact form of the manifold. This result does not rely on the particular case explored in this manuscript and constitutes a general result for 3-dimensional electromagnetic fields. That is, a material medium described by a metric constitutive relation is a superconductor whenever the manifold is a 3-dimensional $\varepsilon$-contact metric structure. Motivated by this 
we show that, in general, every $\varepsilon$-contact 3-manifold that is also K-contact and $\eta$-Einstein is a vacuum solution to NMG as long as the coupling constants are restricted by \eqref{lambdam}.

Furthermore, we provide an explicit example by considering a Hopf fibered $S^3$ as a contact manifold with a \emph{ Lorentzian} metric satisfying our desiderata. We show that, in addition of being a vacuum solution of NMG, the geometric structure allows one to couple the theory to the Maxwell-Chern-Simons action and the Abelian Higgs model respectively. Furthermore, to the best of our knowledge, this is a new vacuum configuration for New Massive Gravity and when coupled to the Maxwell-Chern-Simons action,
we  obtained a new charged solution to NMG-MCS.
In every case the fields were found to be rigid, completely fixed by the couplings. Moreover, all but one of the coupling constants
was found to be free in each case. To us, this indicates a rather high degree of naturalness.
The solutions with matter content where found to
smoothly connect with the vacuum case in the appropriate limits. We also showed that the pair $(S^3,g)$ as a 3-dimensional spacetime is a Petrov D, Kundt spacetime. 
 
 These results appear to be natural in the sense that we considered  $S^3$ as the Hopf-fibration, where the Reeb vector field associated with the contact form is, in fact, a Beltrami field. Thus, it is not surprising that it corresponds to a solution to the helicity integral of the field, which provides a measure of the degree in which the field lines are linked (cf. Chapter 5 in \cite{otway2015backlund}). The helicity integral is, in turn, defined in terms of the Hodge dual of the metric \eqref{metric}, which is itself associated with the contact structure. Therefore, as one might have expected, the self gravitating solutions are completely determined from the value of the cosmological constant $\Lambda$ or, equivalently, from the penetration depth $\ell$ in the corresponding analogue superconducting material.

This analysis has shown us that metric contact manifolds might play an important role in the exploration of 3-dimensional field theories. Moreover, it has opened new ways to understand 3-dimensional superconductors in terms of Beltrami fields on a $\varepsilon$-contact manifold. This constitutes a completely geometric picture of the macroscopic  phenomenon of superconductivity  which may shed some light on its higher dimensional counterpart.

\section*{Acknowledgements}

Discussions with Eloy Ay\'on-Beato and Fabrizio Canfora are gratefully acknowledged. We thank the anonymous referee for insightful suggestions.
DFA would like to thank the Mexican Secretariat of Public Education (Secretar\'ia de Educaci\'on P\'ublica) for support under grant PRODEP 12313509.

\appendix
\bibliography{Heis}

\end{document}